\documentclass[prb,twocolumn,preprintnumbers,amsmath,amssymb,floats,citeautoscript,nobalancelastpage]{revtex4-1}
\usepackage{graphicx}% Include figure files
\usepackage{dcolumn}% Align table columns on decimal point
\usepackage{bm}
\usepackage{txfonts}

\begin{document}

\title{Photo-Seebeck Effect in ZnO}

\author{Ryuji~Okazaki$^{1\ast}$}
\author{Ayaka~Horikawa$^1$}
\author{Yukio~Yasui$^{1,2}$}
\author{Ichiro~Terasaki$^1$}

\affiliation{$^1$Department of Physics, Nagoya University, Nagoya 464-8602, Japan}
\affiliation{$^2$Department of Physics, Meiji University, Kawasaki 214-8571, Japan}

\date{\today}

\begin{abstract}
We examine how the photo-induced carriers contribute the thermoelectric transport, 
i.e. the nature of the photo-Seebeck effect, 
in the wide-gap oxide semiconductor ZnO for the first time.
We measure the electrical conductivity and the Seebeck coefficient with illuminating light.
The light illumination considerably changes the Seebeck coefficient as well as the conductivity, 
which is sensitive to the photon energy of the illuminated light.
By using a simple parallel-circuit model, we evaluate the contributions of the photo-induced carriers to
the conductivity and the Seebeck coefficient,
whose relationship shows a remarkable resemblance to that in doped semiconductors.
Our results also demonstrate that the light illumination increases both the carrier concentration and the mobility,
which can be compared with impurity-doping case for ZnO.
Future prospects for thermoelectrics using light are discussed.
\end{abstract}

\maketitle

\section{Introduction}

Thermoelectric material is a functional material that converts heat into electricity and vice versa through the Seebeck and Peltier effects \cite{TE}.
High-efficiency thermoelectric material is of particular importance because of its potential application for energy-conversion technology using  waste heat \cite{Mahan}.
Although the thermoelectric conversion efficiency is insufficient at present, 
various important concepts to improve the efficiency have been recently proposed \cite{TO}.

The fundamental route for efficient thermoelectrics is to increase 
the dimensionless figure of merit $ZT \equiv S^2T\sigma/\kappa$,
where $S$, $\sigma$, $\kappa$, and $T$ are the Seebeck coefficient, electrical conductivity, thermal conductivity, and absolute temperature, respectively.
In conventional semiconductors, 
the $S^2\sigma$ value (power factor) is maximized with optimal carrier concentration of the order of 10$^{19}$ cm$^{-3}$ 
owing to a trade-off relation between $S$ and $\sigma$ in terms of carrier concentration \cite{Mahan},
and the thermal conductivity is mainly governed by the lattice contribution $\kappa_L$ near room temperature.
Thus, optimization of the carrier concentration is a basic step to improve the thermoelectric efficiency,
which has been typically achieved by chemical substitutions for parent semiconductors.
It was recently shown that an electric-field application to field effect transistor structure on a thermoelectric material
is also an effective tool to control the concentration and improve the thermoelectric property \cite{Pernstich,ohta1,ohta2}.
The induced two-dimensional electron gas exhibits unusually large Seebeck coefficient,
which might be attributed to an enhanced density of states due to the low-dimensional structure \cite{SL,SL1,STO}.

The light illumination may be a simple and useful way to optimize the carrier concentration and improve the effciency.
The conduction phenomenon attributed to the excited carriers by the light illumination is known as the photoconduction.
Likewise, the Seebeck effect contributed from such photo-induced carriers is called ``photo-Seebeck effect'',
which was first measured in conventional semiconductor Ge by Tauc in 1955 \cite{Tauc}.
In 1970s, the photo-Seebeck effect has been explored in several semiconductors Si, GaAs, and CdS \cite{Harper70,Harper70a,Kwok72}.
These materials exhibit finite photo-response for the Seebeck coefficient as well as the conductivity.
The light illumination for Ge increases $\sigma$ and decreases $|S|$, naively understood within the photo-doping effect \cite{Tauc}.
On the other hand, silicon exhibits nontrivial photo-Seebeck effect near room temperature:
Both $\sigma$ and $|S|$ are increased by light illumination, which cannot be understood within a simple model \cite{Harper70}.
The elucidation of the mechanism of such unusual photo-transport behaviors possibly provides a new idea for efficient thermoelectrics.
Note that the photo-Seebeck effect measured with different photon energies might be a powerful tool to investigate detailed electronic band structure as well \cite{Harper70a}.
However, there are only a few reports regarding the photo-Seebeck effect, 
and the thermoelectric transport nature of the photo-induced carriers is poorly understood so far. 

Here, we report  first investigation of the photo-Seebeck effect in the wide-gap oxide semiconductor ZnO (band gap of $E_g\sim3.3$ eV),
which is known as a transparent conducting oxide widely used for applications \cite{zno}.
Aluminum-doped ZnO is investigated as a potential thermoelectric oxide as well \cite{Ohtaki,Ohtaki1}.
We construct the photo-Seebeck measurement system, 
and observe the strong variations of the Seebeck coefficient and the electrical conductivity 
under the ultraviolet light illumination, whose photon energy exceeds the band-gap energy of ZnO.
To analyze the observed data, we utilize a simple parallel circuit model 
and determine the contributions from the photo-induced carriers
existing only near the illuminated sample surface.
Our results demonstrate that the distinct variations of transport properties originate from the changes of
both  carrier concentration and  mobility
with illuminating light,
that is discussed within the framework of conventional semiconductor physics.

\begin{figure}[t]
\begin{center}
\includegraphics[width=0.8\linewidth]{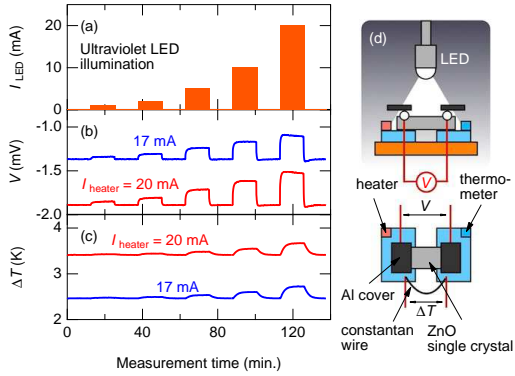}
\caption{(Color online). Measurement time dependence of (a) the excitation current for LED illumination $I_{\rm LED}$, (b) the thermoelectric voltage $V$, and (c) the temperature difference $\Delta T$. The voltage and temperature were measured with several heater-power applications. The illumination was performed with the ultraviolet LED ($\hbar\omega = 3.4$~eV).
(d) The schematic experimental setup for the photo-Seebeck measurement.
}
\end{center}
\end{figure}

\section{Experiment}
We construct the photo-transport measurement system inside the physical property measurement system (PPMS).
The experimental configuration is schematically shown in Fig. 1(d), which is based on ref.~\onlinecite{Tauc}.
We measure the resistivity and the Seebeck coefficient under light illumination with green ($\hbar\omega = 2.4$~eV) and ultraviolet (UV) ($\hbar\omega = 3.4$~eV) light-emitting diodes (LEDs).
The electric contacts are covered with aluminum foils to prevent an extrinsic photo-voltaic effects around such regions.
The electrical resistivity was measured with the two-wire method.
The Seebeck coefficient was measured with standard steady-state method.
The temperature difference $\Delta T$ between hot and cold sides, which is made by applying current $I_{\rm heater}$ to the resistive heater on the hot side,
is determined from the thermoelectric voltage of the bridged constantan wire.
All measurements has been performed at $T = 300$~K.
We used ZnO single crystals (typical dimension of 5$\times$2$\times$0.5 mm$^3$) purchased from a materials company (MTI Corporation).
LED illumination was done to the $ab$ plane.

\section{Results and Discussion}

We firstly describe our experimental procedure for the photo-Seebeck measurements.
We measure the thermoelectric voltage $V$ and the temperature difference $\Delta T$ under illuminations with several different light intensities controlled by the excitation current for LED $I_{\rm LED}$.
Figures 1(a-c) display the measurement time dependence of $I_{\rm LED}$, $V$, and $\Delta T$, respectively.
Here we show the results for the UV LED illumination with five different light intensities.
To eliminate a constant voltage due to a photo-voltaic effect from inhomogeneities inside the crystal, we measure $V$ with several different $\Delta T$ by applying different heater currents $I_{\rm heater}$. 
From these data, we obtain the Seebeck coefficient $S = [V(I_{\rm heater} = 20 \mbox{ mA}) - V(17 \mbox{ mA})]/[\Delta T(I_{\rm heater} = 20 \mbox{ mA}) - \Delta T(17 \mbox{ mA})]$.
As shown in Fig. 1(b), the thermoelectric voltage $V$ is significantly varied with LED light illumination, change of which from the dark value increases with increasing $I_{\rm LED}$. 
We note that the temperature difference $\Delta T$ and the averaged temperature between hot and cold sides are slightly changed by illuminations as shown in Fig. 1(c), however, such  temperature change during the measurement is negligibly small, which cannot explain the observed large $V$ variations under illuminations.
The electrical resistivity is measured using same experimental configuration without thermal gradient.

\begin{figure}[t]
\begin{center}
\includegraphics[width=0.8\linewidth]{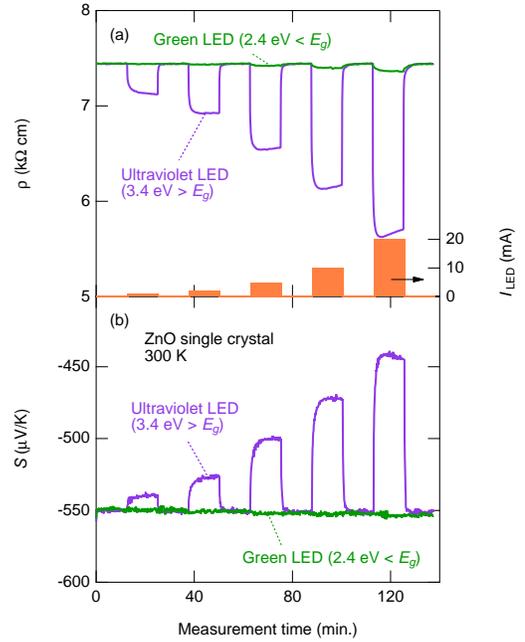}
\caption{(Color online). Measurement time dependence of (a) the electrical resistivity $\rho$ and (b) the Seebeck coefficient $S$ of ZnO single crystal.
The excitation current for LED illumination $I_{\rm LED}$ is also plotted as a function of the time in the right axis of (a).}
\end{center}
\end{figure}

The photo-transport properties of ZnO obtained from above procedures are plotted in Figs. 2(a) and 2(b).
Figure 2(a) shows the resistivity variation measured under five different light intensities indicated as $I_{\rm LED}$ in the right axis.
The UV LED illumination considerably decreases the resistivity, whereas the green light with lower photon energy negligibly affects the resistivity,
indicating that the electron-hole pairs are created by illuminating UV light across the band gap of ZnO and contribute the conduction.
Figure 2(b) displays the Seebeck coefficient measured under illumination.
We observe a significant decrease of $|S|$ with UV illumination, while the green one shows no influence for $S$,
consistent with the photoconductivity results shown in Fig. 2(a).

\begin{figure}[t]
\begin{center}
\includegraphics[width=0.8\linewidth]{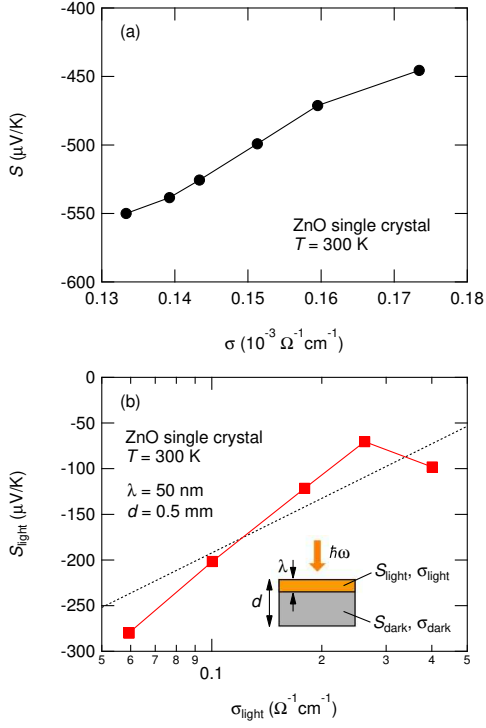}
\caption{(Color online).
Photo-induced transport properties of ZnO under the ultraviolet LED illumination.
(a) The relation between the observed Seebeck coefficient $S$ and electrical conductivity $\sigma$.
(b) The photo-induced Seebeck coefficient $S_{\rm light}$ as a function of the photo-induced electrical conductivity $\sigma_{\rm light}$
(the Jonker plot for the photo-induced transport properties).
The dotted line shows the slope of $k_B/e$.
The inset schematically illustrates our parallel-circuit model composed of light-penetrated and non-penetrated regions.}
\end{center}
\end{figure}

Figure 3(a) depicts the relation between the Seebeck coefficient $S$ and electrical conductivity $\sigma$ measured under the UV LED illumination.
At dark, this sample exhibits an $n$-type conduction ($S$ = -550 $\mu$V/K), which might be attributed to impurity hydrogen \cite{hhh}.
Under UV illuminations, similar to earlier studies on Ge \cite{Tauc}, the conductivity is increased and 
the Seebeck coefficient is decreased in magnitude
as expected in the photo-doping effect.
As discussed in ref.~\onlinecite{Tauc}, this $S$ variation should be related to the difference of mobilities of valence and conduction bands
since the illumination creates electron-hole pairs.
We also note that the 
absorption coefficient of ZnO is about $2\times 10^5$ cm$^{-1}$ at $\hbar\omega$ = 3.4 eV \cite{Yoshikawa97, Srikant98},
indicating that
the illuminated UV light can penetrate into very thin region ($\lambda\sim50$ nm) from the surface.

We then apply a simple parallel-circuit model 
composed of thin conducting and thick insulating layers [the inset of Fig. 3(b)]
in order to analyze the observed photo-transport properties.
The measured electrical conductivity $\sigma$ and Seebeck coefficient $S$ are expressed by \cite{eq_pc}
\begin{equation}
\sigma = \left(1-\frac{\lambda}{d}\right)\sigma_{\rm dark} +\frac{\lambda}{d}\sigma_{\rm light},
\label{sig}
\end{equation}
\begin{equation}
\sigma S =\left(1-\frac{\lambda}{d}\right)\sigma_{\rm dark}S_{\rm dark} +\frac{\lambda}{d}\sigma_{\rm light}S_{\rm light},
\label{sigs}
\end{equation}
where $d$ is the sample thickness (0.5 mm) and $\lambda$ is the  penetration depth (50 nm).
Here $\sigma_{\rm dark}$ ($S_{\rm dark}$) and ${\sigma_{\rm light}}$ (${S_{\rm light}}$) are the electrical conductivities (the Seebeck coefficients) measured at dark and induced by the light illumination, respectively.
Now $d \gg \lambda$, then the photo-induced transport components $\sigma_{\rm light}$ and $S_{\rm light}$ are 
\begin{equation}
\sigma_{\rm light} = \frac{d}{\lambda}\left(\sigma -\sigma_{\rm dark}\right),
\label{sig0last}
\end{equation}
\begin{equation}
S_{\rm light} = \frac{\sigma S-\sigma_{\rm dark}S_{\rm dark}}{\sigma -\sigma_{\rm dark}}.
\label{s0last}
\end{equation}
Figure 3(b) shows the photo-induced Seebeck coefficient $S_{\rm light}$ as a function of the photo-induced electrical conductivity $\sigma_{\rm light}$
obtained from eqs. (\ref{sig0last}) and (\ref{s0last}) (the Jonker plot for the photo-induced transport properties).
The dotted line shows  the slope of $k_B/e$, which is expected in doped semiconductors \cite{Jonker}.
The obtained $S_{\rm light}$-$\sigma_{\rm light}$ behavior roughly coincides with this slope,
implying that the thermoelectric transport property of photo-induced carriers can be discussed in terms of conventional semiconductor physics as well.

\begin{figure}[t]
\begin{center}
\includegraphics[width=0.8\linewidth]{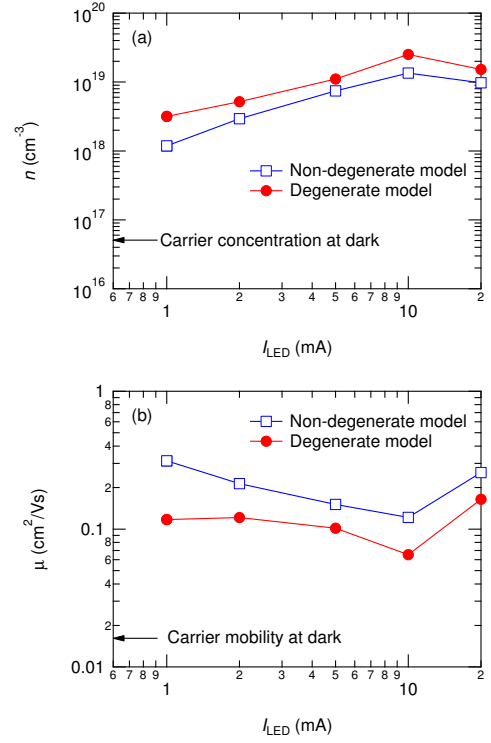}
\caption{(Color online). 
(a) Carrier concentration $n$ and (b) mobility $\mu$ of the photo-induced carriers as a function of the excitation current for LED illumination $I_{\rm LED}$.
Arrows indicate the values measured at dark.}
\end{center}
\end{figure}

We next show the photo-induced carrier concentration $n$, which can be obtained from the photo-induced Seebeck coefficient $S_{\rm light}$.
In the degenerate regime, the Seebeck coefficient is expressed by
\begin{equation}
S = -\frac{\pi^2}{3}\frac{k_B}{e}\frac{k_BT}{\epsilon_F}\left(r+\frac{3}{2}\right),
\label{deg}
\end{equation}
where $\epsilon_F = h^2/2m^*(3n/8\pi)^{2/3}$ is the Fermi energy for a simple parabolic band 
($m^*$ is the effective mass, which is assumed to be $0.3m_0$ \cite{Goano})
and $r$ is a scattering parameter ($r=-1/2$ for an electron-phonon scattering and $3/2$ for an impurity scattering) \cite{thermo}.
In the non-degenerate regime, it is given by
\begin{equation}
S = \frac{k_B}{e}\left[\ln\frac{n}{N_c} - \left(r+\frac{5}{2}\right) \right],
\label{ndeg}
\end{equation}
where $N_c = 2(2\pi m^* k_BT/h^2)^{3/2}$.
Figure 4(a) displays the excitation current $I_{\rm LED}$ dependence of $n$ determined with using eqs. (\ref{deg}) and (\ref{ndeg}).
Here, we adopt $r=-1/2$, which is not crucial to the results.
The carrier concentration at dark is determined with the non-degenerate model [eq. (\ref{ndeg})], 
yielding $5\times10^{16}$ cm$^{-3}$,
as shown by the arrow in Fig. 4(a).
Under illuminations, $n$ is increased from the dark value
in both models, 
and reaches about $10^{19}$ cm$^{-3}$ at maximum light intensity. 
This value corresponds to about 0.1\% Al-doping effect for ZnO \cite{Ohtaki1}
and 
is close to the value of the optimized carrier concentration in conventional thermoelectric materials \cite{TO}.
We have also evaluated the photon number from the output power
 (1.5 mW at $I_{\rm LED}$ = 20 mA) of the UV LED light. 
The number of photons is about $3\times10^{15}$ photons/s. 
However, the electron-hole recombination time in ZnO depends on samples and even a
persistent photoconductivity is observed in several samples owing to
hole traps in grain boundaries and/or oxygen vacancies \cite{Studenikin98,Zhang01}.
Also, we used a wide-angle LED light. These arbitrary properties make it
difficult to quantitatively compare the photo-induced carrier concentration 
with the number of photons absorbed by the sample.

The mobility $\mu$ of the photo-induced carriers can be evaluated as $\mu = \sigma_{\rm light}/en$, shown in Fig. 4(b) as a function of $I_{\rm LED}$.
In contrast to high-mobility ZnO samples \cite{Look},
our sample exhibits very low mobility of
$1.6\times10^{-2}$ cm$^{2}$/Vs at dark,
which is evaluated with using the carrier concentration and the conductivity measured at dark.
The UV illumination considerably increases the mobility roughly by one order of magnitude.
Such mobility variations under illumination have been also reported in photo-Hall measurements on ZnO single crystal \cite{Studenikin},
although they  measured  only transport relaxation process after the illumination.
There, the illumination changes both the carrier concentration and the mobility, consistent with our present results.

Let us discuss more detailed transport mechanism of photo-induced carriers.
In ZnO, the conduction phenomena are mainly governed by polaronic electrons in a broad Zn 4$s$ conduction band,
which are supplied from hydrogenic donor.
On the other hand, the holes in narrow O 2$p$ and Zn 3$d$ valence bands are mostly trapped and negligibly contribute to the transport properties \cite{band}.
Under UV illuminations, therefore, only electrons of the photo-induced electron-hole pairs are mobile and contribute to the photoconductivity and photo-Seebeck effect.
It is also known that the molecular oxygen  in atmosphere significantly affects the transport properties of ZnO \cite{Melnick}.
At the surface, adsorbed oxygen molecules are charged by removing the electrons from the conduction bands, and then 
tightly bound to the surface.
This oxygen adsorption process thus decreases the conductivity.
Under illuminations, such charged oxygen loses its charge by attracting hole of the photo-induced electron-hole pair.
Meanwhile, the electron remains in the conduction bands, leading to an increase of the conductivity.
Such effects are usually dominant in thin films and porous materials with high surface-to-volume ratio, 
but not in single-crystalline bulk materials.
In present case, however, the light-penetrated region is very thin ($\sim 50$ nm), 
then such surface effects should be considered.

Under illuminations, the mobility is also increased
as well as the carrier concentration,
which has been indicated in the photo-Hall experiments as well \cite{Studenikin}.
The origin of the enhanced mobility by illumination is an open question.
In semiconductors, the mobility exhibits a non-monotonic variation as a function of the carrier concentration owing to multiple scattering mechanisms \cite{minami}.
In high-concentration region, where the ionized impurity scattering is dominant, the mobility is decreased with increasing carrier concentration \cite{sca}.
On the other hand, at low carrier concentrations, the mobility is limited by  grain-boundary  or  surface scattering \cite{grain}, 
even for epitaxial films due to a small amount of mosaic structures \cite{epizno}.
This leads to an increase of the mobility with increased carrier concentration,
which might be applied in our present results.
The photo-transport measurement using ZnO samples with different impurity
or vacancy levels such as samples annealed in oxygen or hydrogen is a future issue to elucidate
the exact mechanism of thermoelectric transport property of photo-induced carriers.

We finally give brief comments on the potential thermoelectrics using light.
The light illumination combined with thin film applications may offer a new route toward an efficient thermoelectrics
as well as the fundamental understandings of thermoelectric effects induced by photo-excited carriers.
In bulk measurements, we observe the averaged values including large contributions from non-penetrated region.
On the other hand,
if the sample thickness is smaller than the penetration length (i.e. $d\lesssim\lambda$),
one can directly observe photo-induced components of 
the conductivity  $\sigma_{\rm light}$ and the Seebeck coefficient $S_{\rm light}$.
In present case, the improvement of the mobility with illuminating light directly contributes the $ZT$ enhancement by using thin films.
Illumination on semiconductors with large absorption coefficient may produce
an unexpectedly large photo-Seebeck effect owing to its low dimensionality.
Furthermore, light is able to be utilized as a heat source to make a thermal gradient as well, then
a light-concentration-type thermoelectric module is 
possibly an efficient photo-thermoelectrics using thin films.

\section{Summary}
We have investigated the photo-Seebeck effect as well as the photoconductivity in the wide-gap oxide semiconductor ZnO
for the first time.
We observe a significant variation of such transport quantities, which are sensitive to the photon energy of illuminated light.
The ultraviolet light illumination decreases both the absolute value of Seebeck coefficient and the resistivity, 
acting as a photo-doping effect to ZnO.
A parallel-circuit model is applied to extract the photo-induced contributions from the observed transport data,
which is discussed through the Jonker analysis.
Our results demonstrate that both the carrier concentration and the mobility are increased from dark by illuminations, 
which can be qualitatively argued within the framework of conventional semiconductor physics.
Thin film applications may be efficient for photo-thermoelectrics.

This work was supported by Advanced Low Carbon Research and Development Program (ALCA) from JST
and Program for Leading Graduate Schools ``Integrative Graduate Education and Research Program in Green Natural Sciences'' from MEXT, Japan.

\end{document}